# An introduction to radar Automatic Target Recognition (ATR) technology in ground-based radar systems

Jiangkun Gong, Jun Yan*, Deyong Kong, and Deren Li

*Abstract*— This paper presents a brief examination of Automatic Target Recognition (ATR) technology within ground-based radar systems. It offers a lucid comprehension of the ATR concept, delves into its historical milestones, and categorizes ATR methods according to different scattering regions. By incorporating ATR solutions into radar systems, this study demonstrates the expansion of radar detection ranges and the enhancement of tracking capabilities, leading to superior situational awareness. Drawing insights from the Russo-Ukrainian War, the paper highlights three pressing radar applications that urgently necessitate ATR technology: detecting stealth aircraft, countering small drones, and implementing anti-jamming measures. Anticipating the next wave of radar ATR research, the study predicts a surge in cognitive radar and machine learning (ML)-driven algorithms. These emerging methodologies aspire to confront challenges associated with system adaptation, real-time recognition, and environmental adaptability. Ultimately, ATR stands poised to revolutionize conventional radar systems, ushering in an era of 4D sensing capabilities.

*Index Terms*—Automatic Target Recognition (ATR), ground-based radar systems, 4D radar.

## I. THE DEFINITION OF RADAR ATR

For nearly a century, radar technology has been a cornerstone of detection and ranging systems [1][2]. Radar encompasses a multi-step signal processing chain that involves detection, classification, and tracking. Among these, echo classification stands as a pivotal process. Traditionally, human operators have relied on signal amplitude, trace patterns, and auditory cues, as illustrated in Fig.1, to classify signals, a process distinct from automatic target recognition (ATR).

Radar ATR, in essence, represents an algorithmic or systemic approach to recognizing signals within cluttered backgrounds, utilizing signal processing techniques and pattern modeling methods [3]. To clarify the distinction, NATO experts delineate "target acquisition" as "The detection, identification, and location of a target in sufficient detail to permit the effective employment of weapons," while "recognition" refers to determining the nature, class, or type of a detected entity [4], in accordance with the NATO AAP-6 Glossary of Terms and Definitions revised in 2008. This underscores the marked disparity between radar "detection" and "recognition," with an accent on machine-driven algorithms.

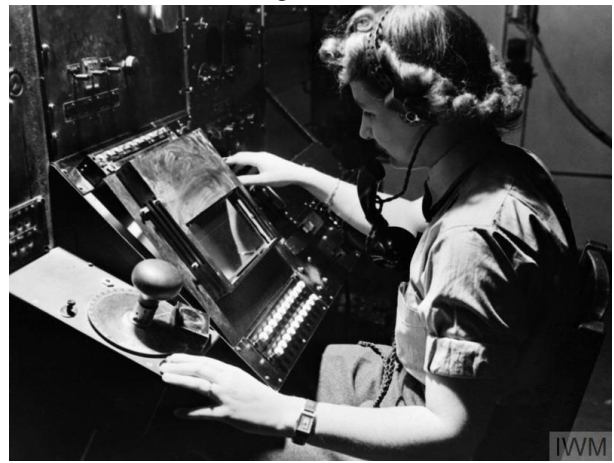

**Fig. 1** WAAF radar operator Denise Miley plotting aircraft on the CRT (cathode ray tube) of an RF7 Receiver in the Receiver Room at Bawdsey CH[1].

1.   This image is cited from https://www.iwm.org.uk/history/how-radar-changed-the-second-world-war.

In surveillance scenarios, radar often detects numerous objects, generating multiple "blips" on the radar screen. The ATR function is then tasked with identifying target details behind each blip and providing target attribution, as depicted in Fig. 2. The advantages of radar ATR include enhanced surveillance capabilities, improved target tracking, and a reduced operator workload. By automating the target recognition process, radar ATR systems facilitate faster and more dependable decision-making, especially in intricate and

Jiangkun Gong, Jun Yan, and Deren Li are with State Key State Key Laboratory of Information Engineering in Surveying, Mapping and Remote Sensing, Wuhan University, No. 129 Luoyu Road, Wuhan, China (e-mail: gjk@whu.edu.cn, yanjun_pla@263.net, drli@whu.edu.cn).

Deyong Kong is with School of Information and Communication Engineering, Hubei Economic Research Room, No.8 Yangqiaohu Road, Wuhan, China (e-mail: kdykong@hbue.edu.cn)

*Corresponding author: Jun Yan (yanjun_pla@whu.edu.cn, +86-027-68778527)

swiftly changing situations. In contemporary military and surveillance applications, the swift and precise identification of targets remains pivotal for situational awareness and decision-making, positioning radar ATR as a valued gem in the crown of radar technology.

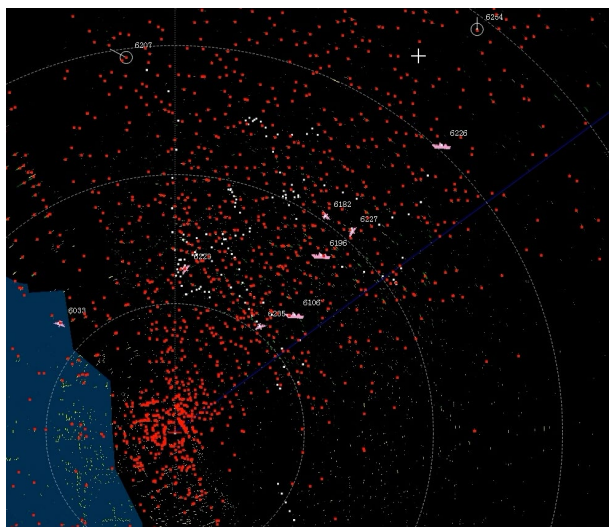

**(a)**

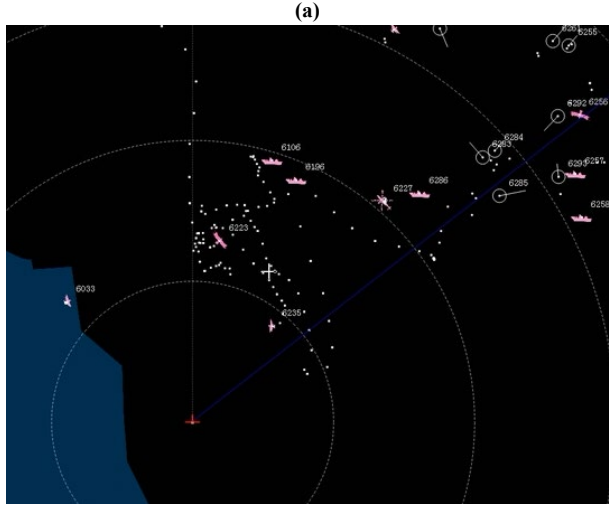

**(b)**

**Fig. 2** The ATR explain the blips on the radar screen[1], **(a)** the raw radar PPI image containing plenty of clutters blips, **(b)** the processed radar PPI image containing only targets.

1. These screenshots are extracted from a real radar screen, which was monitoring East Sea China, around Qigong city, China. More details can be found on our early projects [5][6].

## II. THE MILLSTONES IN THE RADAR ATR HISTORY

Radar ATR technology has been under investigation for over seven decades. One of the earliest radar systems capable of identifying the type of target was likely the SDS ("Système de Détection et de Surveillance" in French, means "Detection and Surveillance System" in English), designed in 1955, a product of Thomson-CSF (France, Now Thales), along with its modification such as RATAC (Fig. 3a) and RASIT. These radar systems were significant innovations at the time as they allowed operators to differentiate between various types of targets, like personnel, vehicles, and aircraft by utilizing the Doppler-frequency audio signal, which was then interpreted by individuals wearing headphones [7]. However, it's important to note that this method wasn't fully automatic and still required human intervention for audio signal classification. Despite these advancements, several challenges remain in realizing practical ATR approaches. In this section, we will summarize significant milestones and provide an overview of the brief history of ATR technology, as listed in Table 1.

TABLE 1 Some milestones in radar ATR history

| Time | Events |
| --- | --- |
| 2016 | The U.S. DARPA conducted ATR oriented research into the TRACE program. |
| 2010 | Our group introduced the first CWS technology for the ATR radar, specifically the ST-312 |
| 2006 | Our Group pioneered the development of the first ATR radar in China, known as BAL-XXX. |
| 2000 | Thales developed the first ATR radar, BOR-A-550, in the world. |
| 1995 | Dr. Chen introduced the micro-Doppler theory. |
| 1990s | The U.S. conducted investigations into the renowned MSTAR and SHARP projects. |
| 1980s | Radar ATR technology is recognized as one of the 20 key U.S. defense technologies. |

The origins of radar ATR development can be traced back to the early years of radar technology, with its initial applications dating back to World War II. In the 1940s and 1950s, researchers embarked on the exploration of methods for the automated detection and classification of targets using radar signals. Substantial progress occurred during the 1960s and 1970s within the realm of pattern recognition techniques for ATR. Researchers devised algorithms that facilitated the comparison of radar echoes with stored target templates or patterns, greatly advancing target identification and classification capabilities.

The 1980s and 1990s are commonly referred to as the inaugural "golden age" of ATR research, partly spurred by the United States' recognition of radar ATR technology as one of the 20 key defense technologies in 1986. During this era, ATR research prominently focused on the extraction of meaningful features from radar data and the application of advanced classification algorithms. Conventional techniques, such as Fourier analysis, wavelet analysis, and statistical pattern recognition, were harnessed to enhance target identification accuracy. However, it's worth noting that the HRRP technology-driven ATR method faced limitations in addressing sample-related challenges, restricting its applicability.

The period spanning from 1990 to 2010 is often referred to as the " silent decade" as ATR technology remained somewhat distanced from practical applications. The 2000s saw a surge in the adoption of machine learning techniques, especially neural networks, within radar ATR, along with the development of micro-Doppler theory. Researchers began utilizing machine learning algorithms to extract micro-Doppler signatures for ATR applications.

Post-2010, the introduction of deep learning techniques, notably convolutional neural networks (CNNs), Deep Learning (DL), and Big Models, ignited a revolution in radar ATR. CNNs, celebrated for their exceptional success in image recognition, had a profound impact on radar ATR technology, ushering in a second "silver age" marked by significantly improved target recognition accuracy. Importantly, this development predominantly centered around the use of radar images, including synthetic aperture radar (SAR) and micro-Doppler spectrograms.

These significant milestones underscore the collaborative efforts of numerous researchers in advancing radar ATR technology over the years. It's crucial to acknowledge that ATR research is a collective endeavor, drawing upon the expertise of scientists, engineers, and institutions from across the globe. The field remains dynamic, continuously evolving, thanks to the ongoing contributions of researchers in academia, government research laboratories, and industry sectors.

(1)      In 1958, the first application of ATR concept was conducted by Dr. Barton identified the corner reflector structure on the Spark II using AN/FPS-16 radar [8][9]. It ushered the born of radar ATR technology, and marked the start of the radar ATR technology.

(2)      Early in 1990s, maybe in 1998, the U.S. DARPA/Air Force Research Laboratory (AFRL) proposed the Moving and Stationary Target Acquisition and Recognition (MSTAR) program is developing state-of-the-art model based vision approach to Synthetic Aperture Radar (SAR) Automatic Target Recognition (ATR) [10], and the SHARP (Systems-Oriented High Range Resolution (HRR) Automatic Recognition Program) projects [11][12][13]. MSTAR is a project aiming at developing model-based ATR technology. A small part of this databased of MSTAR was then released to the public at the beginning of the 2000s, and has motivated a substantial body of SAR ATR research [14]. The data comprise military targets imaged over 360° of aspect angle in spotlight mode and approximately 30 km$^2$ of natural and cultural clutter gathered in strip-map mode [4]. This is the beginning of the famous HRRP technology.

(3)      Around 2000s, maybe in 1995, Dr. Chen proposed the micro-Doppler theory [15]. Micro-Doppler refers to the small variations in the Doppler frequency of a radar return caused by the motion of components or parts of a target. It provides additional information beyond the traditional Doppler effect, which is primarily influenced by the overall motion of the target. The micro-Doppler theory and the corresponding methods has been popular since then, in the radar ATR field.

(4)      The Germany BOR-A-550 radar [16] may be the first one in the world which claimed to be equipped with the ATR function in 2000. They advertised that the BOR-A-550 radar can ATR objects including pedestrians, vehicles, helicopters and ships (Fig. 3b). It is said that the Germans used machining learning technology to fulfil the ATR function.

(5)      In the meantime, in 2006, our group from Wuhan University, developed the first ATR radar, BAL-XXX in China, maybe the second ones after BOL-A-550. It can identify radar echoes from pedestrian, light vehicles, tank, helicopters. And then, in 2010, our group further propose the Classify-While-Scan (CWS) Technology [5] to enhance Situation Awareness power, in the radar, ST-312 (Fig. 3c).

(6)      Nowadays in the last ten years, thanks to the rapid development of deep learning (DL) technology, both America and China have proposed projects from time to time to support the ATR researches. Due to the classified reason, many projects cannot be available for public. Yet, we can also search some of them in media report. For example, In 2016, DARPA proposed the Target Recognition and Adaption in Contested Environments (TRACE) program seeks to develop an accurate, real-time, low-power target recognition system that can be co-located with the radar to provide responsive long-range targeting for tactical airborne surveillance and strike applications.

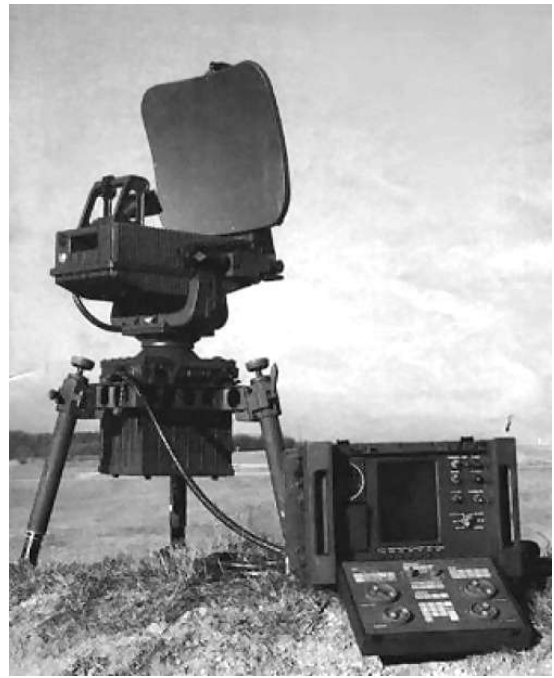

(a)

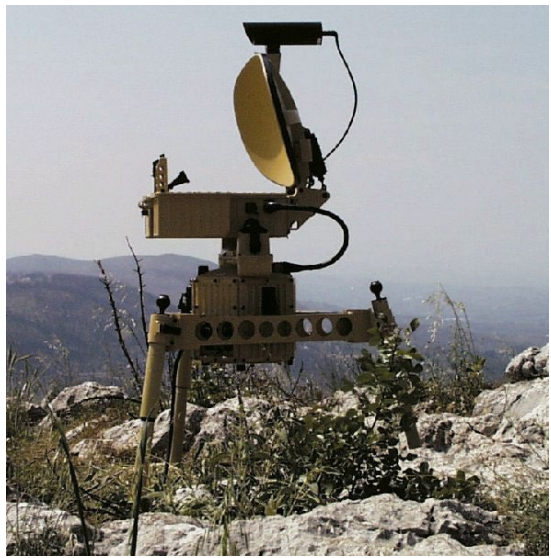

**(b)**

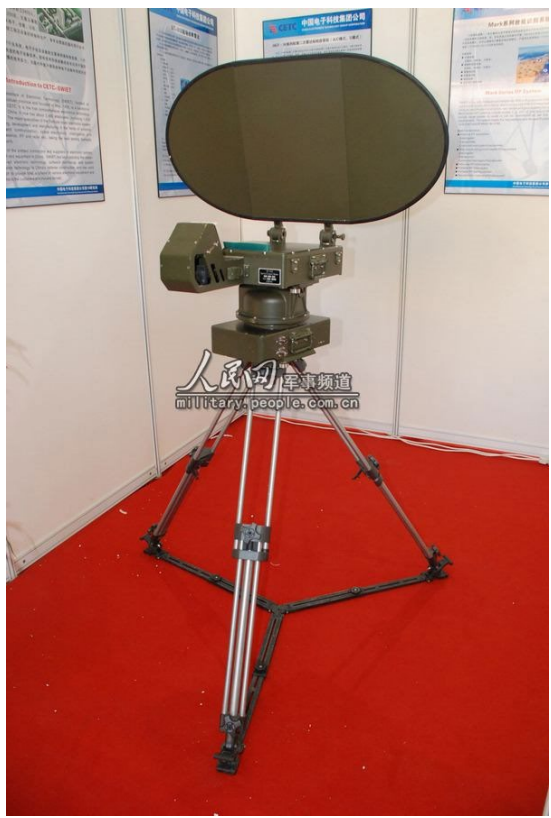

**(c)**

**Fig. 3** The photo of referenced radars, **(a)** RATAC radar[1], **(b)** BOR-A-550 radar[2], **(c)** ST-312 radar[3]

1. This image is cited from https://www.radartutorial.eu/19.kartei/11.ancient2/karte054.en.html
2. This image is cited from https://www.radartutorial.eu/19.kartei/05.perimeter/karte004.en.html
3. This image is cited from https://www.mwrf.net/news/cn/2009/260.html

The classic ATR algorithms use the traditional model-based procedure, including the "feature extraction" plus "pattern recognition". However, the recent ATR algorithms using deep learning (DL) methods look more like a black box testing, which is driven by data. Current efforts in radar ATR focus on real-time processing and integration with embedded systems. The goal is to enable rapid target identification and decision-making, particularly in time-critical applications such as military surveillance and autonomous systems.

## III. ATR FUNDAMENTALS

### A: Recognition tiers

The NATO AAP-6 Glossary of Terms and Definitions, last revised in 2008, offers a comprehensive framework for comprehending ATR and assessing target attributes. Target acquisition is defined as the process involving the detection, identification, and precise location of a target, with the aim of enabling effective weapon utilization. Conversely, recognition pertains to the determination of the nature, class, type, and, when feasible, sub-class of a detected person, object, or phenomenon. Within the realm of ATR, the term "recognition" holds a more specific connotation as per the NATO AAP-6 Glossary of Terms and Definitions [4]. This process can be envisioned as a classification tree, wherein targets are systematically categorized into progressively refined sub-classes as one traverses through the hierarchical structure. Six major classification steps are outlined:

(1) **Detection:** Separating targets from other objects in the scene.

(2) **Classification**: Assigning a meta-class to the target, such as aircraft or wheeled vehicle.

(3) **Recognition**: Specifying the class of the target, such as fighter aircraft or truck.

(4) **Identification**: Assigning the sub-class of the target, such as MIG29 fighter aircraft or T72 tank.

(5) **Characterization**: Considering class variants, such as MIG29 PL or T72 tank without fuel barrels.

(6) **Fingerprinting**: Conducting a more precise technical analysis, such as MIG29 PL with a reconnaissance pod.

It is essential to acknowledge that the boundaries between these decomposition steps may not be universally rigid for all problems and targets. In the research community, there often exists a lack of precision in delineating the performance tiers of ATR, with the terms "classification" and "identification" sometimes used informally when discussing the subject. This lack of clarity can result in confusion regarding the effectiveness of radar solutions. Moreover, relying solely on a single radar capability may not yield higher-tier capabilities, such as "Fingerprinting." Therefore, for radar ATR Tier, we propose a revision to the NATO recognition tier as follows:

(1) **Detection:** Extracting the radar signals of targets in the scene.

(2) **Classification**: Assigning a meta-class to the target, such as drone or bird.

(3) **Identification**: Assigning the sub-class of the target, such as fixed-wing drone or quad-rotor drone.

(4) **Description**: Conducting a more precise technical analysis, such as quad-rotor drone, DJI Phantom 4, with a payload of Go-Pro camera.

### B: Scattering regions theory

Generally, the ATR function is typically seen as an additional module integrated into an existing radar system. This integration implies that the ATR module operates within the confines of the radar's predetermined parameters. Consequently, the ATR module's performance may be constrained by specific radar parameters that may not be ideally suited for the particular ATR solution. Among these parameters, one of the frequently overlooked aspects is the radar band, or more precisely, the associated scattering regions, as illustrated in Fig. 4. A target's scattering behavior can be classified into three distinct regions based on the ratio of its size to the radar wavelength [17]: the Rayleigh region, the resonance region, and the optical region.

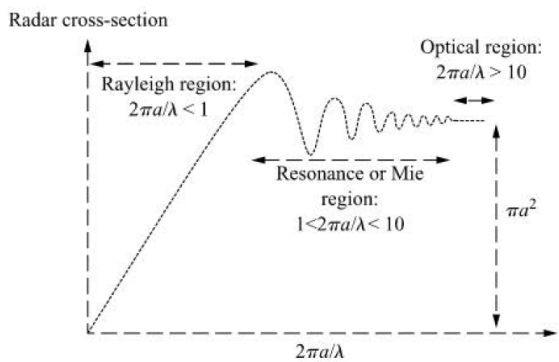

**Fig. 4** The variation of RCS of spheres within three scattering regions, where $a$ is the radius of the sphere, $\lambda$ is the wavelength [18][19]

(1) Rayleigh region

Within this region, the target's size perpendicular to the wavefront is significantly smaller than the wavelength of the incident wave. Consequently, the incident wave does not undergo substantial phase changes when interacting with the target. The RCS of the target follows a linear relationship with frequency and can be approximated as a point source. Echo data obtained from this region provides only rudimentary information such as target size and volume.

(2) Resonance region

In the resonance region, the target's size perpendicular to the wavefront is comparable to the incident wavelength. The scattering behavior of the target primarily arises from surface waves. The RCS of the target becomes a function of both the target size and the wavelength, leading to echo characteristics that exhibit alternating peaks and valleys due to interference between the scattering field components. The resonant region unveils the inherent frequency structure of the target, and a radar system employing multiple polarization states can provide a comprehensive description of the target's poles. Analyzing these poles enables determination of the natural frequency, and then the material composition and identification of the target type. Thereby, "pole" theory related to the natural frequency is basic in the resonance region.

(3) Optical region

Within the optical region, the target's size perpendicular to the wavefront greatly exceeds the incident wavelength. The RCS of the target tends to remain relatively constant. Scattering primarily occurs through specular reflection and the field of the edge segment, which are determined by the strong scattering points present on the illuminated surface of the target. The total strength of the scattering field can be approximated as the sum of these strong scattering points. Radar echoes received from this region encompass detailed geometric and structural information of the target, rendering them valuable for ATR purposes. "Scattering centers" theory related to the geometry is the basic in the optical region.

Different scattering regions necessitate the consideration of distinct ATR methods. Notably, Peter Trait has contributed significantly to the field of radar ATR and has authored a book addressing this subject, which outlines fundamental principles governing radar ATR solutions [17]. Similarly, David Blacknell and Hugh Griffiths have published a comprehensive book on radar ATR [4], delving into various aspects of target categorization, encompassing ground targets, air targets, and maritime targets. Over the course of several decades, multiple ATR schools have recognized the significance of radar features in achieving successful ATR applications in specific cases.

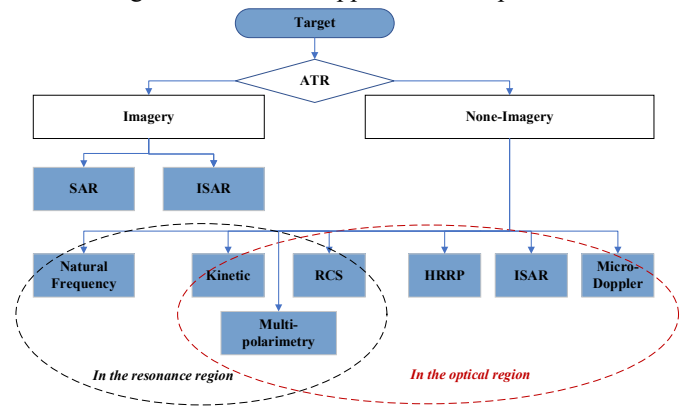

**Fig. 5.** The flowchart of designing a radar driving by different ATR methods

## IV. ATR METHODOLOGIES

Radar ATR technology encompasses both image and non-image data (Fig. 5). Radar imagery includes Synthetic Aperture Radar (SAR) and Inverse Synthetic Aperture Radar (ISAR) images, with SAR being prevalent in air-based radar systems and ISAR commonly used in sea-based radar systems. However, in the context of our discussion, we are focused on ground-based radar systems, where image data is not employed. Instead, we rely on non-image radar signatures. These radar signatures provide features that can be broadly categorized into two distinct groups: long-term characteristics and transient ones. It's important to note that any ATR methodologies discussed should be considered within the context of the scattering region.

### A: Using natural frequency in the resonance region

In the resonance region, the scattering pole is intricately linked to a target's natural frequency, determined by its geometry and material composition. Because the natural frequency remains unaffected by changes in range and altitude, it offers a robust characteristic that can significantly enhance target recognition [20][21][22][23]. It is crucial to emphasize that harnessing this method necessitates a foundational understanding of scattering polar theory within the resonance region, including the concept of " scattering poles." Nevertheless, extracting the natural frequency and associating it with specific objects poses a considerable challenge, resulting in practical radar systems employing this approach being rare in the market. Most research in this domain continues to be confined to laboratory settings due to the ongoing complexity surrounding the scattering mechanisms in the resonance region.

However, it's worth noting that in this resonance region, the Radar Cross Section (RCS) of a target can be significantly amplified. This effect has spurred the development of anti-stealth radar systems that leverage this theory. Anti-stealth radar specifically refers to radar systems engineered to detect and track stealth aircraft or other elusive targets with reduced radar signatures. Given that most aircraft are typically on the order of meters in size, anti-stealth radar systems typically operate in the meter-bands, such as UHF frequencies. Moreover, the scattering pole related to the natural frequency of an aircraft has potential utility in ATR. Nevertheless, due to the highly classified nature of anti-stealth applications, few details are publicly disclosed. However, it is worth mentioning that the fundamental theory underlying these applications often centers around Mie scattering principles.

### B: Using range profile in the optical region

Geometry information is more like extracted in the optic region. Scattering centers of a target in the optic regions can be mapped in many profiles, and then the projected profiles are the model of the target, which is used for matching and recognition. There are two ways to separate the scattering centers and obtain the projected profiles, including the range profile, and the cross-range profile.

The High Range Resolution Profile (HRRP) technology is very first idea that is been used to separate the scattering centers of a target [24][25][26]. Radar systems transmit ultrawide-band signals and capture target profiles in the range direction. The HRRP provides a representation of the target's shape (Fig. 6a), which is then processed using template matching techniques within the dataset to determine the target class. The U.S. Air Force Research Laboratory (AFRL) conducted a notable project called the Systems-Oriented High Range Resolution (HRR) Automatic Recognition Program (SHARP) [13]. In essence, the

HRRP maps the target's geometry information projected in the range domain, which is defined by the high range resolution equation, given by

$$R_e = \frac{c}{2B} \tag{1}$$

where, $R_e$= the range resolution, $c$= the transmitted velocity of light, and $B$= the transmitted bandwidth. Initially, the challenge with HRRP lay in the difficulty and cost of transmitting ultrawide-band signals. However, with the advancement of the global electrical industry and the burgeoning growth of the electrical motor industry since 2016, the radar industry as a whole, have seen significant improvements.

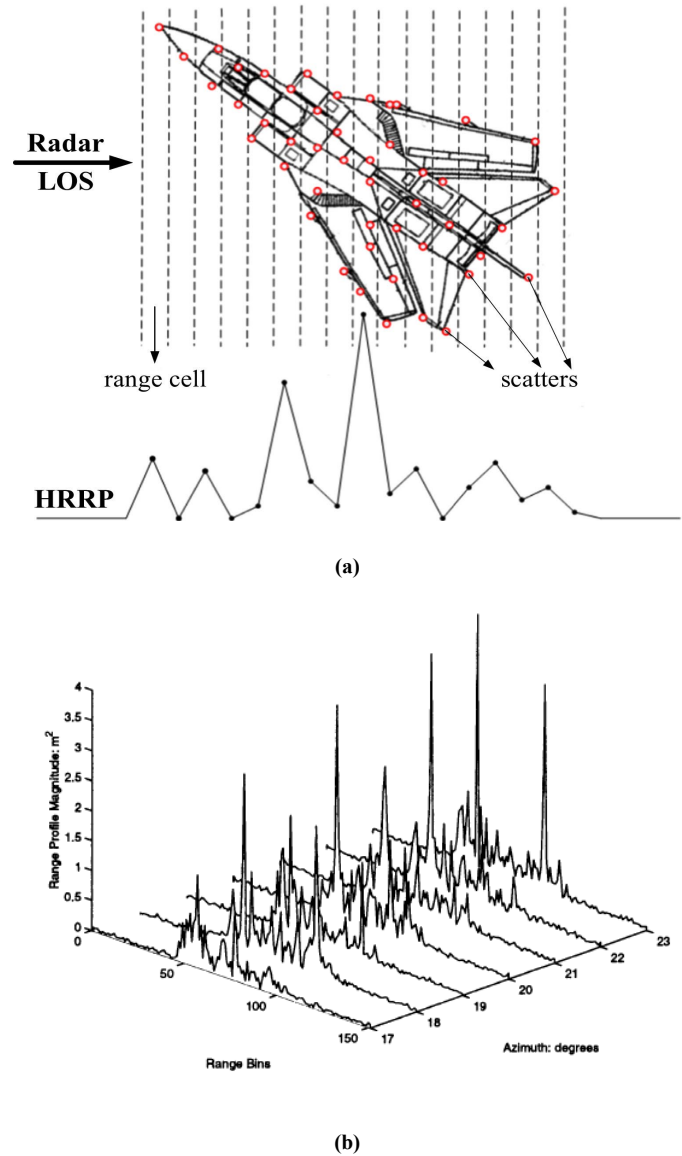

**Fig. 6.** The principle and the attitude-sensitivity problem of the HRRP technology, **(a)** Radar returns from the scatterers on the target are projected onto the LOS, resulting in an HRRP [24][25], (b) Simulated L-band range-profiles of a fire truck for 4 ground vehicles over 6 deg sweep in azimuth [26]

Today, the hardware for HRRP appears to be less of a core barrier, while the algorithm remains a challenge in the field. Apart from the common issue of range cell migration, the primary challenge associated with HRRP technology lies in its sensitivity to a target's orientation. Radar HRRP data typically exhibit high-dimensional, non-Gaussian, and inter-dimensionally dependent distributions, posing a challenging task for parametric modeling in HRRP-based target recognition. As radar angles change, different HRRPs are obtained, as demonstrated in Fig. 6b. Reducing the attitude-sensitivity of HRRP is key to addressing this challenge. A general approach is to measure a wide range of target angles for HRRP, such as at 0.5-degree intervals, and store them in the dataset. This dataset can then facilitate more accurate model matching. However, implementing this idea in practical short-to-medium-range surveillance radar systems remains difficult. Consequently, HRRP may find utility primarily in very costly, large-scale, ultra-long-range surveillance radar systems, such as the U.S. SBX anti-missile radar.

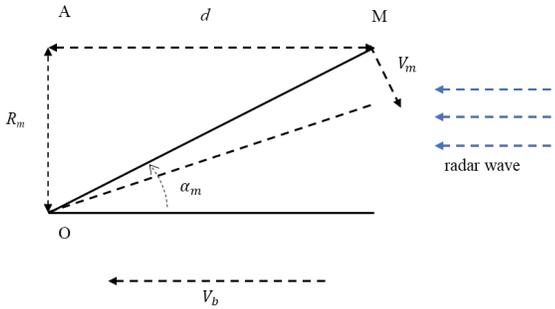

**Fig. 7.** The geometry of the radar and the scattering centres of a target

### C: Using cross-range profile in the optical region

High Doppler resolution technology also serves as a means to distinguish the scattering centers of a target. In theory, since Doppler resolution is directly linked to cross-range resolution, the cross-range profile of an object effectively separates the target's scattering centers. As shown in Fig. 7, assume the scattering centers of a target are $M$, $O$. And the scattering center $M$ rotates around the center of $O$, with the rate of $\omega$. If we neglect the body move speed, $V_b$; and then, and the high cross-range formular is described by

$$\Delta f_{md} = \frac{2L_m \sin(\alpha_m)\omega}{\lambda} \tag{2}$$

where, $\Delta f_d =$ the Doppler resolution, $L_m =$ the length of the two scatter centers, $\omega =$ the angular rotation rate, $\alpha_m =$ the rotating phase angle, and $\lambda =$ the radar wavelength. One of the most renowned applications of cross-range technology is Inverse Synthetic Aperture Radar (ISAR) technology [18]. Cross-range resolution improves with shorter wavelengths and larger rotation angles. Notably, cross-range resolution remains independent of the target's range, distinguishing it from Synthetic Aperture Radar (SAR), where the synthetic aperture must be extended for maintaining range resolution at longer ranges.

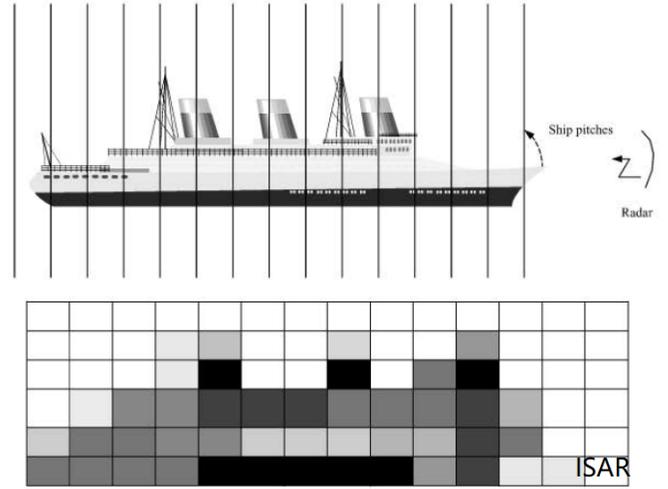

**Fig. 8.** the representation of ISAR image of a ship target after radar data initially collected in each high-resolution down-range gate [18].

Indeed, micro-Doppler also facilitates the separation of scattering centers and can be derived using the high cross-range equation (6). This calculation takes into consideration the body's movement speed, denoted as $V_b$, which is determined through Doppler measurements within the radar system. The Doppler shift can be expressed as:

$$\overline{f_{bd}} = -\frac{2V_b}{\lambda} \tag{3}$$

where, $\overline{f_{bd}} =$ the Doppler shift, and $\lambda =$ the radar wavelength. And, we treat the scattering centers in equation (6) as the micro-scattering, and we introduce the real time rotating phase, $\alpha_m = \omega t$, and then micro-Doppler shift can be given as,

$$\overline{f_{md}}(t) = \frac{2L_m \sin(\omega t)\omega}{\lambda} + \overline{f_{bd}} \tag{4}$$

According to this equation, the micro-Doppler shift undergoes modulation by two sinusoidal functions. Consequently, a comprehensive depiction of micro-Doppler signatures incorporates a modulation function that inherently relies on the observation time. While Dr. Chen, regarded as the pioneer of micro-Doppler theory, derived it using the micro-kinetic model [15], we present a comprehensive overview of micro-Doppler theory from the perspective of cross-range theory. This perspective provides a different angle but one that aligns more naturally with the phenomenon.

Micro-Doppler analysis serves as a valuable tool for extracting embedded kinematic and structural information, essential for target characterization. However, varying theoretical perspectives can result in different processing methods. When considering micro-Doppler from a kinematic perspective, the general approach involves observing time-varying features through integration over adequate radar dwell time and with a high sample frequency [15]. Conversely, within our context, we adopt the cross-range concept for micro-Doppler processing. This approach utilizes differentiation to perceive micro-Doppler signatures as scattering features with the ability to extract structural and geometric characteristics. These two signal processing approaches are further elaborated in.

$$f = \begin{cases} \iint \overline{[f_{md}(t)]} \\ d\overline{[f_{md}(t)]} \end{cases} \tag{5}$$

Fig. 9 illustrates two manifestations of the micro-Doppler phenomenon in radar signals of same type of quad-rotor drones: Jet Engine Modulation (JEM) or Helicopter Rotor Modulation (HERM), which manifests as modulation lines in the spectrum (Fig. 9a), using $d\overline{[f_{md}(t)]}$; the 'blade flash' patterned spectrogram, obtained through the Short-time Fourier Transform (STFT) (Fig. 9b), using $\iint \overline{[f_{md}(t)]}$. JEM spectrum refers to spectral peaks characterized by specific adjacent intervals and similar amplitudes in the spectrum [27], while the 'blade flash' pattern describes the sinusoidal traces in the spectrogram. The instantaneous micro-Doppler signature captures the stable mapping of scattering centers on rotating blades in the cross-range profile, revealing the scattering characteristics or structures of the rotating component rather than the rotational pattern.

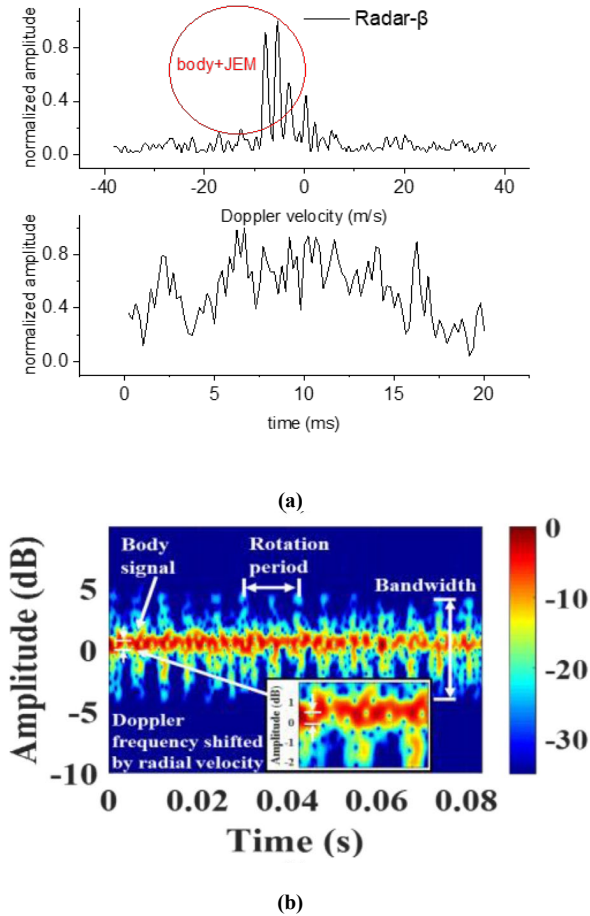

**(a)**

**(b)**

**Fig. 9.** The prestation forms of micro-Doppler of quad-rotor drones, **(a)** the JEM-like spectrums detected by a X-band pulse-Doppler radar [31], **(b)** the "blade flash" patterned spectrogram detected by a X-band radar [32].

While Micro-Doppler signals are influenced by radar dwell time [28], frequency resolution [29], and signal-to-noise ratio [30], it's important to note that the micro-Doppler method exhibits a limited detection and recognition range but entails a relatively lengthy detection response time (DRT). However, despite these inherent challenges, micro-Doppler recognition remains an integral component of contemporary radar systems, facilitating the accurate identification and classification of diminutive aerial targets like drones. As the utilization of drones continues to rise, the importance of micro-Doppler recognition in radar detection is expected to expand significantly.

Our emphasis on the cross-range profile in the optic region is rooted in the potential interference between resonance Doppler frequencies and motion Doppler frequencies when conducting cross-range profiling in the Mie region. Such interference may give rise to indistinct areas around scattering centers, leading to challenges in identifying the source of the cross-range profile and resulting in profile blurriness. Therefore, we advocate for conducting cross-range profiling in the optic region as it offers improved clarity and accuracy in this regard.

### D: The statical features across scattering regions

In many traditional radar ATR applications, static values, such as Radar Cross Section (RCS), speed, and trajectory, are commonly employed as features. Radar systems are capable of detecting targets and measuring various attributes associated with them, which can be categorized into two main groups: RCS-related values and kinetic values. RCS, denoting the target's scattering power or electrical size, serves as a ubiquitous signature for classification. The RCS of a target, denoted as σ, can be expressed as follows:

$$\sigma = \lim_{r \to \infty} 4\pi r^2 \frac{E_s^2}{E_i^2} \tag{6}$$

where, $r$ = the range, $E_s$ = the scattered field, and $E_i$ =the incident field. RCS, or Radar Cross Section, represents the area intercepting the same amount of power as if it were radiated isotropically and produced the equivalent received power in the radar. It is a static mean value. For example, as depicted in Fig. 10, a jet typically exhibits an RCS level of 100 m², whereas a small quad-rotor drone's RCS hovers around 0.1 m². Notably, the measured RCS can exhibit fluctuations over time, often represented as a waveform or time series. Furthermore, statistical features derived from RCS values can yield valuable insights into the target, including altitude, which can be beneficial in target recognition.

Recently, multi-polarimetry/dual-polarimetry has garnered new attention in ATR field. In contrast to conventional single-polarization radar systems, which operate exclusively with either horizontal or vertical polarizations, dual-polarimetric radar simultaneously employs both horizontal and vertical polarizations. This multi-polarization approach yields a more comprehensive and information-rich dataset, rendering it a valuable tool across a diverse range of applications. Particular emphasis has been placed on exploring the dynamics of differential reflectivity (ZDR) within target profiles. However, current multi-polarimetric radar data primarily focus on processing large-scale targets, such as meteorological phenomena [33], biological group targets [34], and chaff [35], with radar systems predominantly designed for weather-related applications. In the context of ATR, the utilization of multi-polarimetric data revolves around identifying patterns within the polarimetric domain of targets' RCS. This analysis primarily relies on static features. To potentially enhance the utility of multi-polarimetric radar in ATR, it may be worthwhile to explore the resonance region of radar data, which could provide more robust insights for ATR applications.

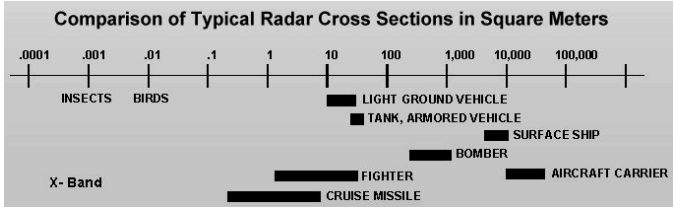

**Fig. 10.** the typical RCS tables within X-band [36]

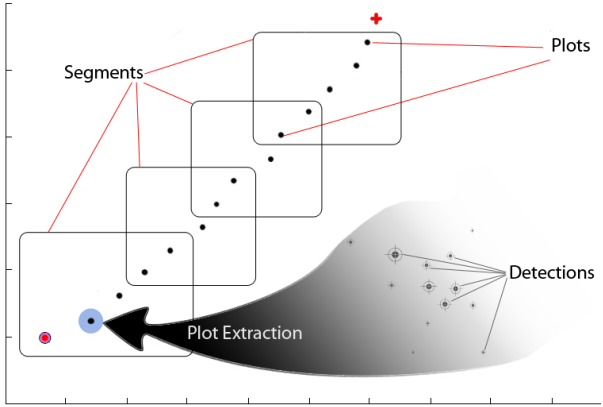

**Fig. 11.** The example of radar trace classification [38]

The kinetic features are highly related to the speed. If the moving speed of the target's body is still the $V_b$, subsequently, the trace function of the target, denoted as $L(t)$ can be defined as:

$$L(t) = V_b t \qquad (7)$$

where, $t$= the measured time. The fundamental kinetic features encompass differential attributes derived from the observed trajectory, taking into account factors such as velocity, time, and range. These features, often referred to as trace classification, have long been a staple in ATR history, especially for distinguishing and categorizing aerial targets. When a radar system detects an aerial target, it generates a trace, essentially a chronological record of the target's movements. Analyzing these traces through trace classification algorithms facilitates the determination of various target characteristics, including size, speed, and flight pattern [37], as illustrated in Fig. 11. Such information plays a pivotal role in classifying the target as a drone, airplane, helicopter, or other types of aerial vehicles. Yet, trace classification demands extended processing time and has a limited recognition range, owing to the less distinctive and robust nature of kinetic features.

## V. ATR VALUES

ATR not only excels in recognizing radar signals but also proves advantageous in radar detection and tracking processes. The origin of radar technology lies in "Radio Detection and Ranging (RADAR)", with its initial task being detection. Through continuous detection, we can establish the trajectories of objects, eventually discerning the target of interest amidst clutter. However, when ATR functions effectively, it enables the separation of the radar detection unit from the tracking unit. In the detection unit, objects are extracted and identified by the ATR module, subsequently providing us with the means to track the target utilizing its identification in the tracking unit. This segregation of functions enhances the efficiency and accuracy of the radar system as a whole. In essence, ATR plays a pivotal role in reducing both "False alarms" and "Missed alarms" without requiring direct target tracking.

### A: Extending Radar Detection Range

Traditional radar detection range of a target is given by the radar equation, which is limited by the RCS value given a specific radar system. The radar equation is mathematically expressed as follows [39]:

$$R = \sqrt[4]{\frac{P_t G_r G_t \lambda^2 \sigma}{(4\pi)^3 K T_s B_n L(SNR)}} \qquad (8)$$

where $T_s$ = the system noise temperature, $B_n$ = the noise bandwidth of the receiver, $L$ = the total system losses, $K$ = the Boltzmann's constant. $P_t$ = the transmitted power, $G_r$ = the received gain, $G_t$ = the transmitted gain, $R$ = the measured range, $\sigma$ = the RCS of the target, $\lambda$= the radar wavelength. The radar equation underscores that the foundational principles of radar design are rooted in radar detection principles. In simpler terms, smaller targets with lower RCS values will yield lower SNRs, leading to a diminished detection range. Lowering the SNR threshold can introduce a high rate of "False Alarms," while increasing the SNR threshold often presents the challenge of "Missed Targets" for radar systems to contend with.

SNR is employed to describe the detection probability of a target because the extracted signals cannot be readily identified as either the target or clutter. In essence, SNR serves as a preliminary screening method to distinguish potential targets from background clutter. Well-established equations in radar engineering indicate that for single-pulse detection, if the detection probability exceeds 50%, the SNR of the target should be at least 13.1 dB. To achieve a 95% probability of detection, the SNR should be 16.8 dB [39]. However, when the target is irradiated by radar waves, its echoes are always detected by the radar system, and the question is that if can we easily distinguished from the background clutter.

If we treat radar clutter or even noise as potential targets and process radar signals in each radar resolution cell, recognizing them using certain ATR methods, it is referred to as the integrated detection and recognition (IDR) method. This approach allows us to simultaneously reduce both "Missed Targets" and "False Alarms." And, we can extend the radar detection range significantly beyond the theoretically expected ranges, a concept known as beyond-range detection, enhanced by the ATR function. Fig. 12 illustrates an example where radar can detect small birds at distances exceeding 12 km and large birds at distances exceeding 15 km. These detection ranges far surpass the theoretical expectations based on the radar equation. Additional details can be found in our earlier project [40].

### B: Improving Radar Tracking Performance

In addition to radar detection, radar tracking is equally essential and plays a significant role. Radar tracking provides valuable historical information about a target, facilitating the prediction of its future movements—a concept known as situational awareness. Hence, the quality of tracking directly impacts situational awareness. Traditional radar tracking algorithms typically start as detection problems, involving the extraction of an initial target location within a radar resolution cell and the measurement of the target's initial speed for trace prediction. In this context, detection precision and update rate are two critical factors that can both be significantly improved with the assistance of ATR.

Firstly, ATR can enhance detection precision by reducing "False Alarms." Typically, at the beginning of radar tracking, the nature of the object is unknown. However, with the assistance of ATR, we can swiftly identify the target within the surveillance area, significantly expediting the radar tracking process. Secondly, ATR can reduce the reliance on speed related to update rate when predicting the target's next location within the radar tracking algorithm. ATR can shift the tracking strategy from the traditional prediction-based approach to a connection-based strategy. With real-time target identification, we can simply connect traces belonging to the same target, allowing us to obtain the target's trajectory. This approach minimizes the role of speed, especially in scenarios involving multiple targets with overlapping traces—a methodology referred to as the track-after-identify (TAI) method. Fig. 12 illustrates an example where multiple objects, including "small bird 6223," "ship 6266," "helicopter 6227," "ship 6235," "large bird 6273," and "small bird 6283," are monitored in a cluttered sea environment. The ATR function enhances dynamic situational awareness, akin to a "video." Further details can be found in our earlier project [5][40].

### B: Transforming of Radar Functions

Lately, several vendors have introduced the concept of 4D radar. Some argue that "4D imaging radar" represents a high-resolution, long-range sensor technology that offers significant advantages over 3D radar, particularly in determining object height. Others claim that 4D radar encompasses the ability to measure a target's range, azimuth, height, and velocity. Nevertheless, traditional radar systems can determine the 3D location and speed of a target through techniques like Doppler processing or image differencing.

The term "4D radar" is often used as a marketing buzzword and does not precisely correspond to the fourth dimension in physics. From our perspective, a traditional radar primarily measures the 3D location, and the fourth dimension in physics pertains to the object's characteristics. In this context, a radar system can indeed be considered a 4D radar, as it not only captures the spatial dimensions (azimuth, elevation angle, and slant range) but also incorporates information about the target's inherent attributes, as the one in Fig. 12.

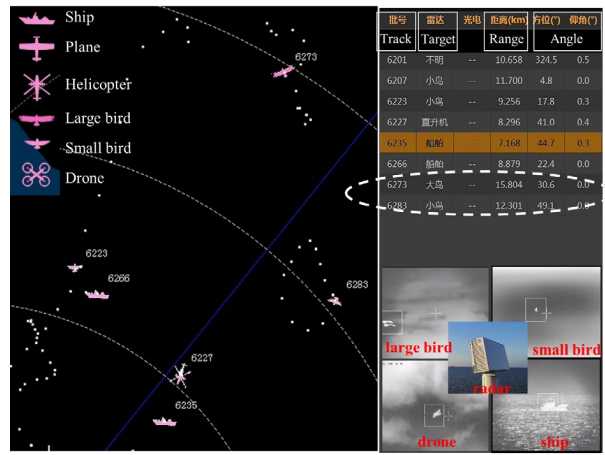

**Fig. 12.** Example showing that small radar can detect birds at the range over 12 km, and provide the situational awareness using CWS technology [40],

## VI. ATR APPLICATIONS IN URGENT WAY

The ongoing Russo-Ukrainian War has highlighted the urgent need for ATR (Automatic Target Recognition) technology in addressing three critical applications. On February 24, 2022, Russia launched a comprehensive invasion of Ukraine, marking a modern warfare scenario. However, it is noteworthy that many contemporary radar systems deployed in the conflict zone have exhibited suboptimal performance. Notable examples include the famous S-400 air defense system, as well as various ship-based radars in the Russian Black Sea fleet. These systems have faced challenges posed by small drones, stealth missiles, loitering missiles, and other lightweight weaponry, leading to their degradation or destruction. The lessons learned from this war emphasize the crucial importance of advancing technologies to address three primary areas of concern: countering drone threats, improving anti-jamming techniques, and enhancing anti-stealth capabilities.

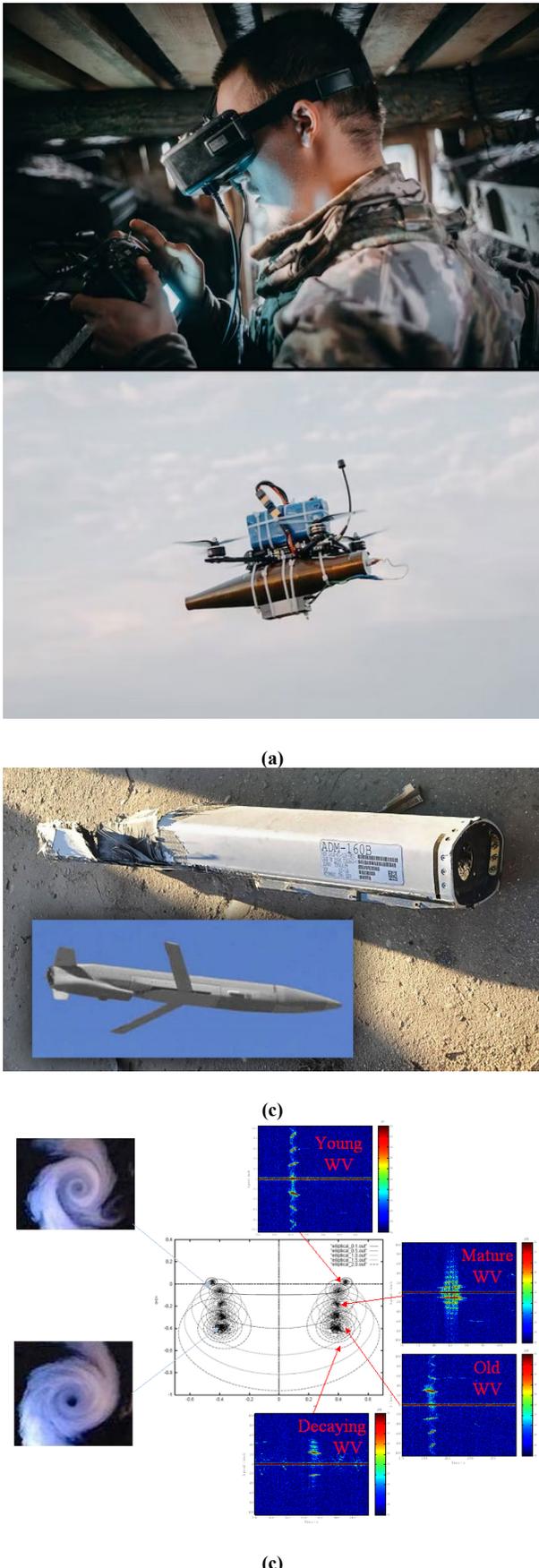

**(a)**

**(c)**

**(c)**

**Fig. 13.** The three ATR applications, **(a)** Drone operator flies an FPV drone from a forward bunker position on the southern front line near Robotyne, Ukraine, on Sept. 14. [1], **(b)** Evidence Of ADM-160 Miniature Air-Launched

Decoy Use By Ukraine Emerges[2], **(c)** Evolution of Roll-up spiral geometry & Doppler spectrum(time/Doppler slopes) versus Age [42][41].

1.  This image is cited from https://www.washingtonpost.com/world/2023/10/04/fpv-drone-ukraine-russia/(Wojciech Grzedzinski for The Washington Post)
2.  This image is cited from https://www.thedrive.com/the-war-zone/evidence-of-adm-160-miniature-air-launched-decoy-use-by-ukraine-emerges

## A: Countering drone threats

The first critical lesson arises from the threat posed by small drones. Small drones typically fall into the category of low-speed, low-altitude, and low-RCS targets, often referred to as LSS targets in radar detection terminology. These LSS drones typically operate at speeds below 200 km/h, at low altitudes below 1000 meters, and possess a small RCS ranging from 0.1 to 1 $m^2$. Traditional air surveillance radars, designed primarily to detect large, fast-moving objects such as helicopters, jets, and missiles, often struggle to differentiate between LSS drones and birds due to their similar radar signatures. To illustrate this challenge, consider Fig. 13a, which depicts an FPV drone equipped with missiles in a forward bunker position on the southern front line near Robotyne, Ukraine. These FPV drones, akin to birds in their radar profiles, can serve as agile weapons for repelling invading forces. Consequently, the detection and tracking of LSS drones pose significant difficulties for traditional air surveillance radars, given the intermittent traces they leave.

The primary technological challenge in the context of Counter-Unmanned Aerial System (C-UAS) radar applications lies in the classification of radar echoes emanating from both birds and drones, as they exhibit radar signatures that are often indistinguishable from each other. Additionally, there is a need to identify the specific type of drone encountered, as different drone types require distinct countermeasures. For instance, small quad-rotor DJI Phantom drones might be effectively countered using lasers, while large fixed-wing TB2 drones may necessitate missile-based interception. It is worth noting that the conventional ATR method relying on trace recognition is not an ideal solution in C-UAS radar applications, as trace segments can sometimes be elusive or unavailable for analysis. Consequently, the development of dedicated ATR methods tailored to counter-drone applications becomes imperative. In summary, the utilization of drones presents air defense systems with a novel and intricate challenge, emphasizing the critical need to develop effective counter-drone capabilities to ensure security and safety.

## B: Anti-Jamming (AJ) Applications

The second lesson is derived from the realm of jamming technology. Jamming methods are extensively employed by both offensive and defensive forces. Jamming entails transmitting signals to disrupt the normal functioning of a radar system, potentially leading to the failure of the radar system and even impacting the outcome of a battle. In military and defense applications, the reliability of radar systems is paramount, emphasizing the critical need for anti-jamming techniques. As shown in Fig. 13b, Ukraine may have employed ADM-160 Miniature Air-Launched Decoys during this conflict. Depending on the specific variant, these decoys can be utilized to jam enemy radars or deceive radar operators, creating the illusion of threats approaching from various directions. This diversionary tactic often diverts defenders' attention and resources away from actual incoming threats.

The conventional jamming strategy relies on the premise that the signal amplitude of the jamming signal surpasses that of the target. Traditional anti-jamming radar technologies employ agile waveform diversity to counteract the scattering power caused by jamming signals. However, an alternative and more intelligent approach involves the use of Advanced Target Recognition (ATR) functionality, which classifies radar signals from both targets and jamming sources. This method eliminates the need to alter the transmitting waveform and is more adaptive in contemporary radar systems. In summary, implementing ATR-based anti-jamming solutions represents an effective means to protect critical radar systems from deliberate or inadvertent interference, ensuring their reliable operation even in challenging conditions.

### C: Enhancing Anti-stealth Capabilities

The third lesson is learned in the context of stealth aircraft and missiles. While there is no concrete evidence to suggest that either Ukraine or Russia has employed stealth aircraft, there have been instances of stealth missile use, such as the Storm Shadow Missile. For example, on September 22, 2023, Ukraine employed Storm Shadow missiles to attack the Kremlin's Black Sea Fleet headquarters after breaching the air defense radar systems of the Black Sea Fleet base in Crimea. In the contemporary landscape, both stealth aircraft and stealth missiles present challenges for air defense radar systems because they are adept at operating quietly and disruptively. Many stealth aircraft possess exceedingly small Radar Cross Section (RCS) values, sometimes even smaller than those of birds. This characteristic makes them formidable adversaries for traditional air defense radar systems, which often struggle to detect, let alone track, these stealthy threats.

Aircraft and missiles operate in the skies but leave behind a wake vortex, which is essentially a byproduct of their flight. It's important to understand that these wake vortices can also be considered as radar targets, thus making them detectable by radar sensors. For instance, since 2007, Thales, in collaboration with its partners, has conducted a series of experiments using X-band radar and Lidar technology to measure the wake vortex of aircraft at several airports in France (Fig. 13c). These experiments have demonstrated that the X-band Thales BOR-A-550 radar is capable of detecting wake vortices with a Radar

Cross Section (RCS) as low as 0.01 square meters on medium-sized aircraft. This detection capability holds true in all weather conditions, both wet and dry, at short ranges of less than 2 kilometers and even at longer ranges exceeding 7 kilometers. Wake vortices are characterized by their spiral geometry, age decay (youthful, mature, old, decaying), and strength of circulation, as outlined in references [16][41][42].

The wake vortex essentially represents the trail left behind by an aircraft, unveiling various attributes of the aircraft itself, including its speed, type, time since passage, and more. This perspective allows us to regard the wake vortex as an additional element accompanying the primary aircraft. This leads to a thought-provoking idea: What if we harness Advanced Target Recognition (ATR) technology to detect, recognize, and track the aircraft by leveraging the wake vortex? By utilizing ATR technology to detect the wake vortex and subsequently track its lines back to their source, we may have a promising method to track stealth aircraft. This innovative concept may be currently under development in many laboratories.

## VII.  ATR FUTURE

### A: ATR Integration with Cognitive Radar (CR) Platform

Cognitive radar (CR) systems stand as an innovative radar technology specially tailored for the detection of small drones. It introduces a continuous adaptability to evolving sensing tasks by harnessing receiver feedback and employing machine learning techniques to enhance detection performance. Coined by Simon Haykin [43], the term "cognitive radar" encompasses the implementation of four fundamental cognitive features: the perception-action cycle (PAC), memory, attention, and intelligence [44][45][46]. Since its inception, cognitive radar has garnered substantial attention within the radar field, leading to a plethora of studies exploring its capabilities [47][48][49] [50][51]. Essentially, contemporary cognitive radars aim to adapt their transmitted waves to sense environmental clutter, thereby enhancing the detection performance of targets. In essence, cognitive radar can be considered a facet of the ATR process.

One primary challenge of cognitive radar lies in the fact that the variation in transmitted parameters can also influence an ATR method. Hence, it is imperative to investigate this influence, a task that has been undertaken by some researchers. For instance, K. Barth et al. explored a cognitive radar framework for radar classification, utilizing high-range resolution profiles (HRRP) and waveform diversity. They demonstrated that the cognitive waveform approach led to a 15% improvement in classification accuracy compared to a static approach using a single waveform [52]. A. Huizing et al. delved into the potential of incorporating deep learning techniques, such as Convolutional Neural Networks (CNNs) and Recurrent Neural Networks (RNNs), for mini-drone classification using micro-Doppler spectrograms within the context of cognitive radar [30]. Additionally, J. Gong et al. suggested the feasibility of designing a "cognitive micro-Doppler radar" capable of detecting and tracking micro-Doppler signals from drones by adapting the radar dwell time [28]. In summary, cognitive radar has the potential to provide enhanced radar data for ATR performance due to its incorporation of a feedback loop, but it represents a relatively nascent research field with promising avenues for exploration.

### B: ATR solution driven by machine learning (ML) algorithms

Cognitive radar may indeed enhance ATR platform quality, while machine learning algorithms hold the potential to improve ATR algorithm. It's worth noting that we are now entering the third wave of artificial intelligence (AI), known as the 3rd wave of contextual adaptation [53]. This wave follows the 1st wave of handcrafted knowledge and the 2nd wave of machine learning.

In the context of radar ATR, ATR solutions driven by ML algorithms are also advancing rapidly because ATR primarily revolves around classification and recognition tasks. In this era of contextual adaptation, the 3rd wave of AI, future ATR algorithms face several significant challenges: Firstly, they must deliver real-time and on-chip classification capabilities, a critical requirement in many radar ATR applications. For instance, when considering ATR solutions for missile platforms, one must account for strict time constraints imposed by both hardware limitations and the high-speed nature of missiles. Secondly, ATR algorithms should incorporate transfer learning or environment adaptation capabilities. Current ML-based ATR solutions often excel with data from their training datasets but struggle when confronted with unfamiliar datasets, a common issue known as "sample dependency". Finally, and perhaps most critically, ATR algorithms based on ML models should prioritize explain-ability. While advanced ML techniques offer the potential to create ATR solutions capable of autonomous perception, learning, decision-making, and action, they often struggle to articulate the rationale behind their decisions and actions to human users. This poses a significant challenge, particularly for military clients who rely on ATR systems. A completely opaque "black-box" ATR solution, one that cannot shed light on its decision-making process, is simply not a feasible choice for military applications, given the potential consequences of such obscurity. Therefore, ensuring the transparency and explain-ability of ATR systems remains of utmost importance, especially in military contexts.

So, what should the ideal ATR solution look like? In this context, we envision a perfect ATR solution that mimics human visual recognition. Human facial recognition, for example, can occur in as little as 120 milliseconds, with coarser classifications taking just 50 milliseconds [54]. Moreover, humans don't require numerous glances to remember a stranger's face and recognize it from memory. This rapid and efficient recognition process is an ideal benchmark for ATR algorithms.

## VIII. CONCLUSION

In this paper, we present an in-depth exploration of Automatic Target Recognition (ATR) technology in the context of ground-based radar systems. We aim to provide a comprehensive understanding of the ATR concept while tracing the key milestones in its historical development. ATR methods are systematically classified into different scattering regions. The integration of ATR solutions into radar systems holds the potential to extend radar detection ranges and significantly enhance radar tracking capabilities, leading to a substantial improvement in situational awareness. Currently, drawing insights from the Russo-Ukrainian War, there are three pressing radar applications that demand ATR technology: detecting stealth aircraft, countering small drones, and implementing anti-jamming measures. We anticipate that the next wave of radar ATR research will originate from the fields of cognitive radar and machine learning-driven algorithms. These emerging approaches aim to address critical challenges related to system adaptation, real-time recognition, and environmental adaptability. In essence, ATR has the potential to transform traditional radar systems from simple ranging sensors into intelligent sensing centers, ushering in the era of 4D sensing capabilities.

### DECLARATION OF INTERESTS

The authors declare that they have no conflict of interest.

### ACKNOWLEDGMENTS

January 2023 signifies the 50th Anniversary of the IEEE Aerospace and Electronic Systems Society (AESS), a significant milestone in the field. In light of this commemoration, this article serves as a tribute to the radar automatic target recognition (ATR) technology, acknowledging the contributions of radar pioneers throughout history.